# Remote Water-to-air Eavesdropping through Phase-Engineered Impedance Matching Metasurfaces

Jing-jing Liu[1], Zheng-wei Li[1], Bin Liang[1]*, Jian-chun Cheng[1]* and Andrea Alù[2,3]*

[1]*Key Laboratory of Modern Acoustics, MOE, Institute of Acoustics, Department of Physics, Collaborative Innovation Center of Advanced Microstructures, Nanjing University, Nanjing 210093, P. R. China*

[2]*Advanced Science Research Center, City University of New York, New York, NY 10031, USA*

[3]*Graduate Center, City University of New York, New York, NY 10016, USA*

*Correspondence authors. Emails: liangbin@nju.edu.cn (B. L.); jccheng@nju.edu.cn (J.-c. C.); aalu@gc.cuny.edu (A. A.)

**ABSTRACT**

Efficiently receiving underwater sound remotely from air is a long-standing challenge in acoustics hindered by the large impedance mismatch at the water-air interface. Here we introduce and experimentally demonstrate a technique for remote and efficient water-to-air eavesdropping through phase-engineered impedance matching metasurfaces. By judiciously engineering an ultrathin mechanically-rigid boundary, we make the water-air interface acoustically transparent and at the same time we are able to pattern the transmitted wavefront, enabling efficient control over the effective spatial location of a distant airborne sensor such that it can measure underwater signals with large signal-to-noise ratio as if placed close to the physical underwater source. Such airborne eavesdropping of underwater sound is experimentally demonstrated with a measured sensitivity enhancement exceeding 38 dB at 8 kHz. We further demonstrate opportunities for orbital-angular-momentum-multiplexed communications and underwater acoustic communications. Our metasurface opens new avenues for communication and sensing, which may be translated to nano-optics and radio-frequencies.

**TEASER**: Sound-transparent and phase-engineered metasurface for remote water-to-air eavesdropping.

**INTRODUCTION**

What happens under the water-air interface covering 70% of the earth cannot be directly detected through acoustic signals, a challenge that has long been hindering a plethora of applications, such as sonar communications and ocean explorations (*1-6*). Due to the large ratio in acoustic impedance between water and air (around 3600 times), only 0.1% of the acoustic energy can naturally be transmitted through such a nearly-perfect reflective boundary (namely, 30 dB loss), equivalent to the loss experienced in underwater acoustic propagation accumulated over 30 kilometers at 10 kHz (*7*). In comparison to conventional approaches to impedance matching, which feature bulky and impractical devices (*8-10*), recently emerging acoustic metamaterials, such as membrane-type (*11*) and bubble-based (*12-16*) artificial materials, among others (*17-19*), may enable enhanced transmission between water and air in more efficient and compact platforms. However, their application in real-world systems, requiring mechanical stiffness, high endurance and large transmission efficiency, still remains poorly explored. In addition, beyond the transmission loss at the water-air interface, the signal-to-noise ratio (SNR) measured by a remote sensor in air is also substantially affected by diffraction and dissipation effects in the long-range propagation distance, which may be compensated by manipulating and steering appropriately the transmitted signal to enhance the intensity at the sensor location, and thereby achieve high sensitivity. An effective solution to this challenge is elusive at present, since existing metasurface approaches to steer and direct sound waves only work in homogeneous media (*20-26*) or media with weakly-mismatched impedance (*27-29*).

In this article, we propose a technique that can overcome both these fundamental limitations and present the experimental demonstration of remote water-to-air eavesdropping through nearly perfect impedance matching metasurfaces, which simultaneously provides wide-angle total transmission at the water-air interface and arbitrarily mold the spatial sound field with flexibility and efficiency. Our mechanism thereby allows an acoustic sensor in one medium to pick up the sound signal emitted from the other medium with large sensitivity, regardless of the sensor location or the monitoring distance. As a practical implementation, we demonstrate a compact mechanically-rigid structure comprising a resonance matching layer that compensates severely mismatched acoustic reactance induced by near fields to enable near-unity transmission, and a labyrinthine-shaped component effectively controlling at the subwavelength scale the sound transmission path. The experimental realization for airborne eavesdropping of underwater low-frequency signals demonstrates energy redistribution near the sensor and a record-breaking enhanced sensitivity over 38 dB. We also demonstrate how this strategy enables novel applications for orbital-angular-momentum (OAM) multiplexing and underwater acoustics, with fundamental relevance for water-to-air remote communication engineering and underwater stealth.

**RESULTS**

**Operation of the impedance matching metasurface**

Figure 1A schematically shows the common scenario of airborne monitoring of underwater sound. For simplicity, and without losing generality, we assume that sound is emitted from a point source in water, undergoing nearly-total reflection at the bare water-air interface due to the large impedance contrast. This feature, in addition to unavoidable diffraction of propagating waves, leads to extremely low intensities of the measured signal once sensor and source are separated in the two media. An increase in emitted power or reduction in source-sensor distance, as viable solutions to improve the measured SNR, have negligible impact in improving the connection, due to the large inefficiencies.

Our proposed technique to enhance remote water-to-air eavesdropping through impedance matching metasurfaces can drastically enhance the sensitivity at the sensor location, independent of the distance between sensor and source or of the spatial location of the sensor. As illustrated in Fig. 1B, we judiciously decorate the water-air interface with a tailored artificial layer that supports wide-angle total transmission at the otherwise reflective interface, while at the same time enabling arbitrary manipulation of the spatial distribution of the acoustic transmitted energy. This solution compensates for the reduction in signal intensity at the sensor location, and makes it equivalent to moving the sensor to a location very close to the source. By modulating the effective spatial location of the sensor, it is possible to eavesdrop the sound signals with large SNR while keeping a large monitoring distance.

Our proposed acoustic structure for implementing this phenomenon consists of a multi-layered configuration, with an individual unit cell schematically illustrated in Fig. 2A. The proposed unit cell has subwavelength dimensions and a rational design, built by grooving an acoustically-rigid cube with a labyrinthine-like channel filled with air. Such an acoustic architecture overcomes the inherent limitations of existing anti-reflection designs [based for instance on a single layer of bubbles (*12*), hydrophobic materials (*13*) or membranes (*11*)] in terms of low transmission efficiency, short interaction distance and weak mechanical rigidity, as demonstrated in the following. For each unit cell proposed here, layers A and C are coupled to air and water respectively, and the effective acoustic impedance of each layer can be freely adjusted by tuning its air channel width (see Supplementary Materials for design details and geometrical parameters). Layers A and C serve as resonance matching layers that use their strong interaction with the incident wave to enable near-unity transmission. The labyrinthine-shaped layer B, with a length-varying air slit, slows down the sound speed by lengthening the propagation path and thereby introduces a controllable local phase delay to the transmitted wave, patterning at will the transmitted wavefront with subwavelength resolution. We first use numerical simulations to demonstrate the operation of our unit cell: the calculated transmission spectrum, as well as the sound transmission through two particular unit cell designs as shown in Fig. 2B, demonstrates a tunable and linearly continuous phase shift spanning a $2\pi$ range with near-unity transmission efficiency. This enables efficient impedance matching at the water-air interface, enhancing the transmission over 850 times compared to the bare interface, and simultaneously modulating the wavefront on the water-air interface within a subwavelength physical dimension, as required by our mechanism. Notice that the special structure design of layer C significantly increases the effective radiation area, and thereby compensates for the acoustic reactance introduced by the higher-order evanescent modes at the surface of the air-filled channel, which is too large to be simply eliminated by shortening the channel length as in conventional airborne metasurface design (see Supplementary Materials for more theoretical analysis and derivation). To verify this effect, we plot in Fig. 2C the enlarged view of the field distribution inside layer C (marked by the black dashed line in Fig. 2B), with the direction of energy flow at the structure-water interface, from which we can observe the typical intensity distribution of a resonant matching layer and significantly enhanced radiation than the one of a conventional air-filled channel. We also investigate the wide-angle functionality of the device and plot in Fig. 2D the simulated transmission as function of incident angle and length of the labyrinthine path. The results show a high transmission efficiency for a wide range of incident angles, which may be further tuned for the range of angles of interest by tailoring the structural parameters. The resulting artificial matching interface presents a large mechanical stiffness and associated robustness, which are highly desirable for underwater applications. As a feasible implementation, our design makes it possible to effectively and arbitrarily engineer sound waves in media with extremely-mismatched impedance.

**Experimental realization of airborne eavesdropping for underwater sound**

We will verify the effectiveness of our technique by experimentally demonstrating airborne eavesdropping for underwater sound signals. The working frequency is set to 8 kHz (corresponding to a wavelength $\lambda_a$ of 4.3 cm in air and $\lambda_w$ of 18.6 cm in water). In order to modulate the effective spatial location of the airborne sensor and significantly enhance its detection sensitivity, the transmitted spatial sound field needs to be transformed to be directively radiated towards the sensor. In the considered scenario, this is equivalent to focus the otherwise blocked energy on the transmitted side towards the sensor position, which is realized by imposing the phase pattern $\Phi = -ik_a\sqrt{(x-x_0)^2 + (y-y_0)^2 + (x-y_0)^2}$ at water-air boundary, where $k_a$ is the wave number in air and $(x_0, y_0, z_0)$ is the sensor location [set to be $(0, 0, 5\lambda_a)$ in our demonstration]. Based on this principle, we assemble an artificial coupling boundary consisting of 6 × 6 metasurface unit cells to realize near-perfect water-air coupling and apply the desired phase profile at the water-air interface (see Supplementary Materials for more details). Figure 3A shows the photograph of the experimental setup and the prototype fabricated via three-dimensional (3D) printing, respectively (see Methods for detailed experimental setup).

We first measured the sound field distribution over 20 × 20 cm$^2$ square regions in the focal plane and in the vertical plane with and without metasurface, and plotted the results in Fig. 3C. Good agreement is observed between measured and simulated results in Fig. 3B, showing the expected focused energy distribution near the eavesdropping point, in contrast to the dark spot in the case of the bare air-water interface. At the eavesdropping condition, a large sensitivity enhancement over 38 dB is experimentally verified, significantly better than state-of-the-art techniques. For a clear illustration, we also show in Fig. 3D the one-dimensional acoustic pressure profile along the *x* axis in the focal plane, including both numerical and experimental results. These results validate the exceptional performance of our technique for water-to-air eavesdropping through the design of a sound-transparent and phase-engineered metasurface. Our mechanism demonstrates an effective relocation of the sensor in the adjacency of the physical source, thereby enabling eavesdropping with large sensitivity despite the long physical distance. We also explore remote monitoring performance in the far field via numerical simulations, due to the limited size of our experimental setup. The far-field sound pressure level (SPL) is calculated for the cases with and without metasurface in Fig. 3E, showing an average SPL enhancement over angular direction exceeding 25 dB. In addition to the significantly increased sensitivity, omnidirectional acoustic transparency can be observed in the far-field directivity curve, confirming that the impedance matching phenomenon is not limited to a narrow angular range, and paving the way towards various opportunities for acoustic applications.

**Applications of water-to-air eavesdropping**

We demonstrate the generality and applicability of our technique in more complicated eavesdropping scenarios and consider opportunities to shape the acoustic wave in more complicated scenarios. Recent efforts have shown that OAM beams offer a degree of freedom independent of time and frequency to enable significant boosting of channel capacity and transmission speed, in a convenient form of multiple-input-multiple-output communications. However, to date OAM-enabled high-speed acoustic communications has only been explored in homogeneous media (*30-32*). Our metasurface offers the opportunity of properly reshaping the phase profile at the water-air interface, enabling cross-media OAM multiplexing, as demonstrated in Fig. 4 (A and B), and establishing an "acoustic information bridge" linking water and air effectively. From our results, we observe highly-efficient and precise generation of multiplexed OAM beams of desired order, which can serve as independent communication channels through an acoustically impenetrable interface, as evident by the spiral pattern of propagation phase and donut-shaped intensity profile with enhanced amplitude of the transmitted beam. These phenomena overcome the fundamental limitation of existing spatial multiplexing mechanisms that transfer information with significantly-boosted rates but only through a homogeneous medium, offering the possibility for high-speed water-to-air communication and high-security encrypted eavesdropping (*31, 33, 34*).

Our technique can also be applied to eavesdropping underwater sound inside an underwater cavity filled with air. Based on this idea, another example of virtual acoustic window for underwater eavesdropping is explored in Fig. 4 (C and D). The results prove that our approach can convert the total reflection at the air cavity into total transmission, focusing the incident energy towards a predesigned spatial position inside the cavity, and resulting in an enhanced sensitivity of over 42 dB. This effect allows the use of airborne detectors, such as capacitor-type microphones, for underwater high-SNR sound detection, which are generally about 3 orders of magnitude more sensitive than piezoelectric transducers (*11*). Such a "window" effect, along with the much higher absorption of low-frequency sound in air than in water, may open route to exciting opportunities for underwater detection and camouflage (see Supplementary Materials for details on these two applications).

**DISCUSSION**

To conclude, here we have proposed and experimentally demonstrated a technique for remote high-SNR water-to-air eavesdropping using impedance matching metasurfaces. As a practical implementation, a compact and mechanically-rigid artificial coupling unit was judiciously designed to make the water-air interface acoustically transparent over a broad range of incident angles and arbitrarily engineer the transmitted wave, which enables effectively steering the sound wave at interfaces with extremely-mismatched impedance. The performance of our technique was demonstrated both numerically and experimentally via several distinct examples of airborne monitoring of underwater sound signal with an unprecedent sensitivity enhancement and cross-media OAM multiplexing and acoustic window for underwater eavesdropping. The working bandwidth of such metastructure-based artificial boundary is narrow in the present implementation (a Q-factor about 15), but it can be extended by introducing coupled resonances (*14, 35*).Our technique can be also applied to monitor airborne sound signals in the underwater environment according to reciprocity. This suggests the potential of our approach in more complicated scenarios, such as active eavesdropping in which the sensor and source are both located in air to detect the underwater objects. Our methodology provides a promising platform for general water-to-air eavesdropping and it can benefit various applications, such as cross-media communication engineering, remote sensing, underwater vehicle stealthy and beyond.

## MATERIALS AND METHODS

### Simulations

Throughout the paper, the simulations are conducted with a finite element method based on COMSOL MULTIPHYSICS software. The mass density and sound speed of air are set as $\rho_a = 1.21$ kg/m$^3$ and $c_a = 343$ m/s, respectively. The mass density and sound speed of water are set as $\rho_w = 1000$ kg/m$^3$ and $c_w = 1490$ m/s, respectively. The solid material for building the artificial boundary is chosen as UV resin whose mass density and sound speed are 1400 kg/m$^3$ and 2000 m/s respectively.

### Experiments

In the underwater part, a water tank with sound-absorbing rubber mounted on the boundary was built to mimic an anechoic environment. A low-frequency underwater acoustic transducer with central frequency of 8 kHz was placed at the bottom center. The transmitted airborne sound signal was measured in a 3D air anechoic chamber using a 1/4-in. free-field microphone (Brüel & Kæjr type-4961), which was mounted on a 3D stepping motor to scan the target region point by point. The measurements of sound field scanning were taken at every 1 cm in all directions.

**Acknowledgement**

**Funding:** This work was supported by the National Key R&D Program of China (Grant No. 2017YFA0303700), the National Natural Science Foundation of China (Grant Nos. 11634006 and 81127901), the Simons Foundation and a Vannevar Bush Faculty Fellowship, High-Performance Computing Center of Collaborative Innovation Center of Advanced Microstructures and a project funded by the Priority Academic Program Development of Jiangsu Higher Education Institutions. **Author Contributions:** J.L. and B.L. conceived the idea and designed the experiment; J.L. and Z.L. prepared the samples and performed the experiments; J.L. conducted the numerical simulations. J.L., B.L., J.C., and A.A. contributed to the writing of the paper; B.L., J.C., and A.A. supervised the entire study. **Competing Interests**: The authors declare that they have no competing interests. **Data and materials availability:** All data needed to evaluate the conclusions in the paper are present in the paper and/or the Supplementary Materials. Additional data related to this paper may be requested from the authors.

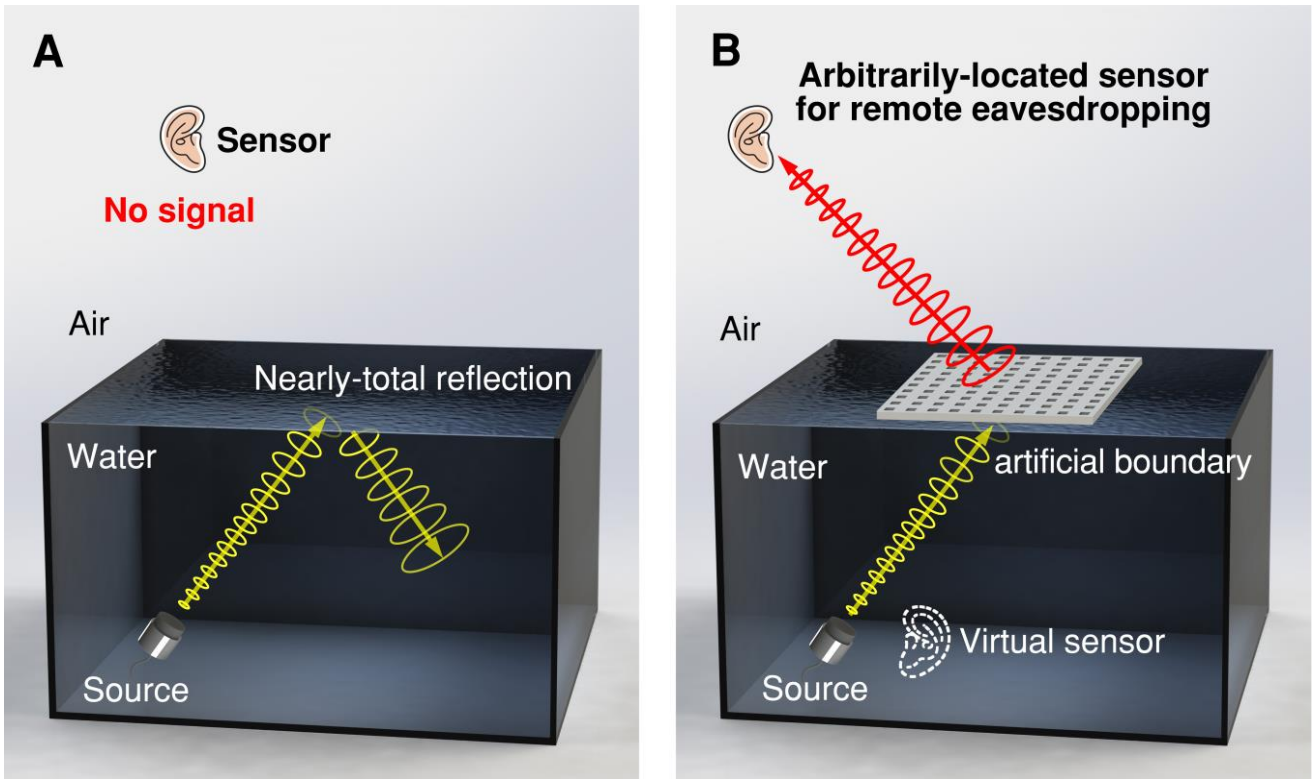

**Fig. 1. Impedance matching metasurface for remote water-to-air eavesdropping.** (**A**) Schematic diagram of airborne monitoring of underwater signals at the water-air interface. (**B**) Proposed principle for remote water-to-air eavesdropping through ultrathin impedance matching metasurface. By judiciously decorating the interface with a tailored artificial boundary, extreme enhancement in sensitivity can be achieved.

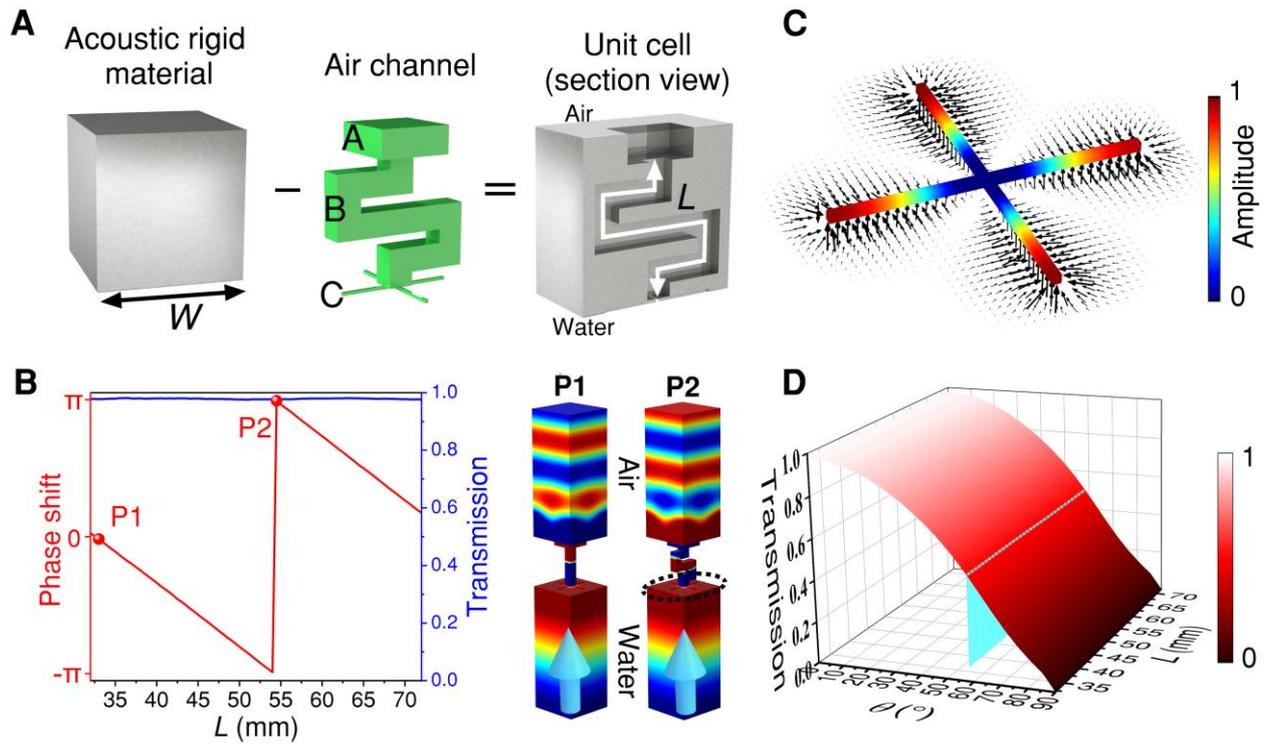

**Fig. 2. Practical implementation of impedance matching metasurface for water-air interfaces.** **(A)** Schematic diagram of a unit cell to impedance match the water-air interface. The side length $W$ equals $0.2\lambda_w$. **(B)** Simulated phase shift and transmission provided by unit cells as a function of $L$. Inset: two specific water-to-air pressure field distributions through designed unit cells (P1 and P2). **(C)** Simulated sound amplitude distribution in layer C [marked by the black dashed in **B**], as well as direction of energy flow at the water-structure interface. **(D)** Simulated transmission as a function of incident angle $\theta$ and structural parameter $L$. The blue line denotes the transmission value of 50 %.

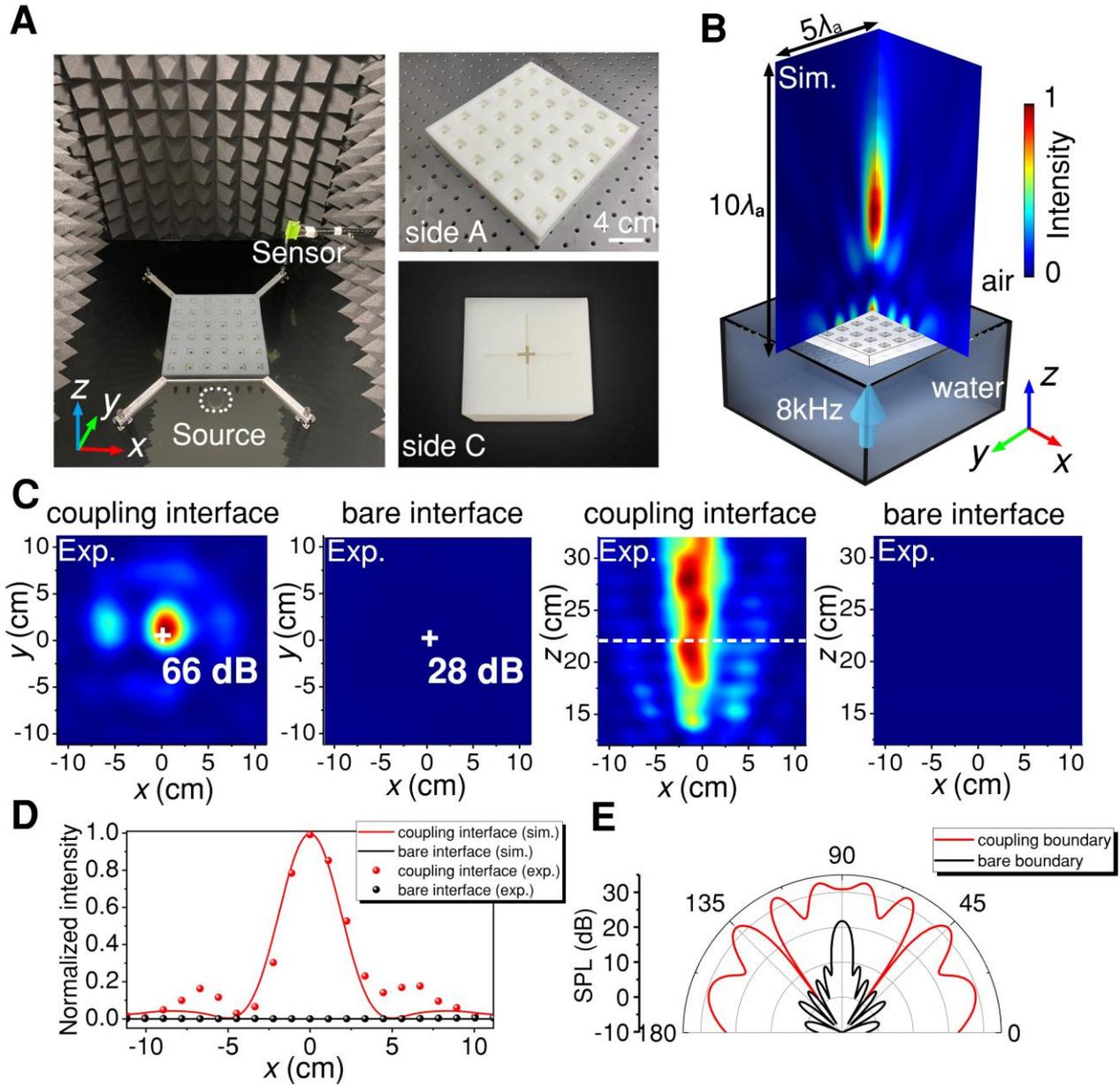

**Fig. 3. Experimental realization of high-SNR airborne eavesdropping for low-frequency underwater signals.** (**A**) Photograph of experimental setup and fabricated sample. The 3D size of the water tank is $0.8 \times 0.8 \times 0.8$ m$^3$. (**B**) Simulated 3D sound intensity distribution of transmitted sound wave in air. (**C**) Measured sound intensity distributions over $20 \times 20$ cm$^2$ square regions in the focal plane and in the vertical plane with and without metasurface. (**D**) Comparison between measured and simulated sound intensity profiles on the axis [white straight line in **C**]. (**E**) Simulated far-field SPL with and without artificial coupling boundary.

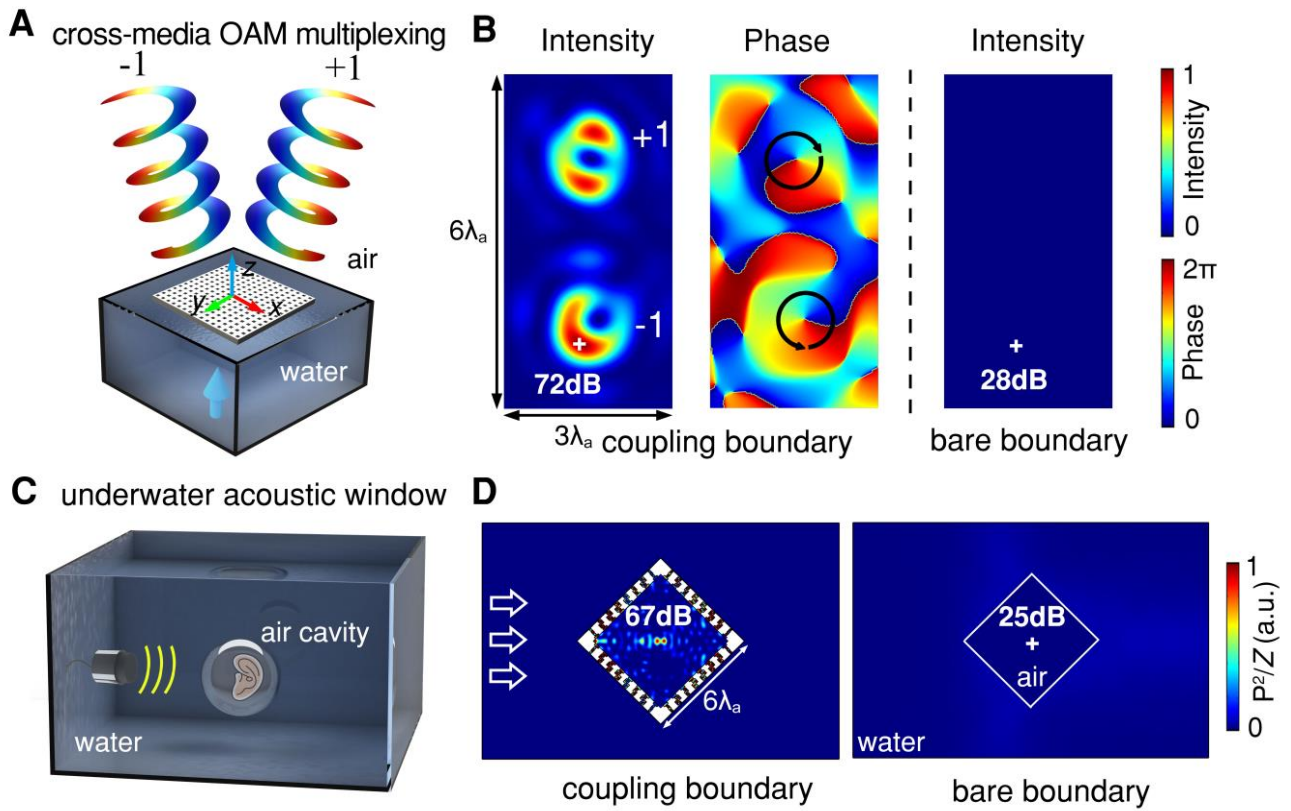

**Fig. 4. Demonstration of cross-media OAM multiplexing and underwater acoustic window. (A)** Schematic of cross-media OAM multiplexing (with orders of 1 and -1 as an example). **(B)** Simulated intensity and phase distributions on the detection plane for OAM multiplexing with and without metasurface. **(C)** Schematic of the acoustic window for underwater eavesdropping. **(D)** Simulated 2D intensity distributions for underwater eavesdropping inside a square air cavity with and without metasurface, respectively. The white arrows denote the incident waves in water.

**SUPPLEMENTARY MATERIALS**

Supplementary materials have been attached to this paper.